\documentclass{ws-procs9x6}

\begin{document}

\title{PROGRESS IN ATOM CHIPS AND THE INTEGRATION OF OPTICAL
MICROCAVITIES}

\author{E. A. HINDS,$^*$ M. TRUPKE, B. DARQUIE, J. GOLDWIN and G. DUTIER}

\address{Center for Cold Matter, Imperial College London, Prince Consort Road,\\
London, SW7 2AZ, UK\\
$^*$E-mail: ab\_ed.hinds@imperial.ac.uk\\
http://www3.imperial.ac.uk/ccm/}

\begin{abstract}
We review recent progress at the Centre for Cold Matter in
developing atom chips. An important advantage of miniaturizing atom
traps on a chip is the possibility of obtaining very tight trapping
structures with the capability of manipulating atoms on the micron
length scale. We recall some of the pros and cons of bringing atoms
close to the chip surface, as is required in order to make small
static structures, and we discuss the relative merits of metallic,
dielectric and superconducting chip surfaces. We point out that the
addition of integrated optical devices on the chip can enhance its
capability through single atom detection and controlled photon
production. Finally, we review the status of integrated
microcavities that have recently been demonstrated at our Centre and
discuss their prospects for future development.

\end{abstract}


\bodymatter

\section{Introduction}

Atom chips offer a powerful way to miniaturize experiments in atomic
physics \cite{fortagh_RMP_2007}. Microstructures on the surface of
the chip produce magnetic and electric fields which can be used to
confine and manipulate cold atoms above the substrate surface.
Experiments have shown that it is possible to transport and split
cold neutral atom clouds or Bose--Einstein condensates in tight
traps near the surface of a chip using either microscopic patterns
of permanent magnetization in a film or microfabricated wire
structures carrying currents or charges. Large field gradients can
be generated close to a microstructured surface, creating tight
traps with oscillation frequencies up to  $\sim 1$\,MHz at the
micron scale where tunneling and coupling between traps become
important. This makes atom chips suitable for applications in
matter-wave interferometry,\cite{Schumm_Nat_Phys_2005} quantum
sensing \cite{Hinds_PRL_2001}, atomic clocks
\cite{Treutlein_PRL_2004}, quantum information processing
\cite{Calarco_PRA_2000}, and the study of low-dimensional quantum
gases \cite{LowDGases_JP4_2004}. However many of these potential
applications require the ability to detect a small number of atoms,
to create versatile microscopic trapping configurations, and to
manipulate the atomic quantum state, both internal and motional.
Optical methods exist to accomplish these tasks, but are only just
now starting to be miniaturized and integrated into atom chips.

One microscopic optical device we are studying is an atom detector
based on a pair of optical fibres facing each other. These single
mode fibres have been machined at their ends, providing small beam
waists at the foci. Light transmitted through one fibre and
collected by the other should allow detection of 5 to 10 atoms
placed in the gap \cite{Eriksson_EPJD_2005}, either through their
absorption (on-resonance imaging) or by the optical phase shift
(dispersive imaging). The fibres can also be used to generate a
one-dimensional standing-wave pattern, making an optical lattice in
which the atoms can be manipulated by the optical dipole force.
However, in order to bring the optical sensitivity down to the
single atom level, we require an optical cavity to increase the
atom-light interaction strength.

The integration of microcavities with the trapping and guiding
capabilities of atom chips opens new possibilities for experiments
in atomic physics. In an optical microresonator, the interaction
strength of an atom with a photon in the cavity mode increases with
decreasing mode volume, as $1/\sqrt{V}$ \cite{Berman_1994}. This is
evident in the expression for the vacuum Rabi frequency in the
cavity,
\begin{equation}
    2g=\mu\sqrt{\frac{\omega_\mathrm{C}}{2\hbar \epsilon_0 V} }\,,
    \label{eq:g}
\end{equation}
where $\mu$ is the dipole moment of the atomic transition and
$\omega_\mathrm{C}$ is the resonance frequency of the cavity. This
interaction is damped by the decay rate of the excited atomic
population, which is $2\gamma$ in free space, and of power in the
cavity, which is $2\kappa=\pi c/(LF)$. Here $L$ is the cavity length
and $F$ is the finesse of the resonator. The driving and damping
rates are succinctly compared in the single atom cooperativity,
$C=g^2/2\kappa\gamma$. The total spontaneous emission rate of an
atom in a cavity can be made to differ greatly from the free-space
value, depending on the cooperativity. With the cavity tuned to the
atomic transition frequency and with $\kappa < g$, the natural decay
rate becomes $2\gamma(1+2C)$. The second term represents emission of
photons into the cavity mode at a rate $4C\gamma=2g^2/\kappa$.

The resonators used successfully in other experiments have reached
single-atom cooperativity values exceeding $50$
\cite{Vahala_Nature_2003}, but are not ideal for atom chip
experiments as they consist of comparatively large components,
making it impossible to bring the intense part of their mode close
to the trapping structures on the chip surface.  For this reason,
research is being carried out on a variety of integrable microcavity
designs. These include microtoroids, fibre-fibre microcavities and
our fibre-chip cavities. Microtoroids offer the highest quality
factors among these, and are well-suited for scalable atom-chip
integration as they can be produced by standard silicon
microfabrication techniques.  Strong coupling between single atoms
and the field of a microtoroid resonator has been
demonstrated\cite{Aoki_Nature_2003} with atoms passing through the
evanescent field in free-fall. To couple light into and out of these
resonators, a tapered fibre is used.  The fibre position needs to be
adjusted so that the evanescent field at the taper overlaps with the
evanescent component of the microtoroid mode.  This is the only
manual procedure required to make the devices operational. While the
strong-coupling condition has been experimentally fulfilled for the
first time for such a device, the challenge of reliably positioning
and trapping atoms in the evanescent field of these devices with the
required accuracy has yet to be surmounted.

By contrast, atoms can be placed directly and accurately into the
region of highest field strength of Fabry--Perot-type resonators.
For this reason, the efforts of several research groups are
currently focussed on this type of resonator. Several experimental
groups have already succeeded in positioning atoms accurately within
the mode of an optical resonator using optical, electrostatic and
magnetic transport techniques. The positioning of a Bose--Einstein
condensate in a microcavity on a chip by means of current-carrying
wire guides has also been recently demonstrated using a
fibre-coupled microcavity\cite{Colombes_quantph_2007}. The
resonators used in that work are however not ideally suited for
scalable integration, as they are constructed in a sequence of
mostly manual steps.

Here we report on recent experiments at our Centre on the detection
of atoms with on-chip micocavities, whose design aims to combine the
advantages of microfabrication and fibre-coupling with the open
access of a Fabry--Perot resonator.

In the next section we describe two typical atom chips that are in
use in our laboratory to magnetically trap atoms either with current
carrying wires or permanent magnet structures.
Section~\ref{section:microcavities} deals with detection of atoms
and enhanced emission of photons using on-chip microcavities. We
conclude with an outlook in Sec.~\ref{section:outlook}.

\section{Magnetic atom chips}\label{section:atom_chips}

\begin{figure}
    \psfig{file=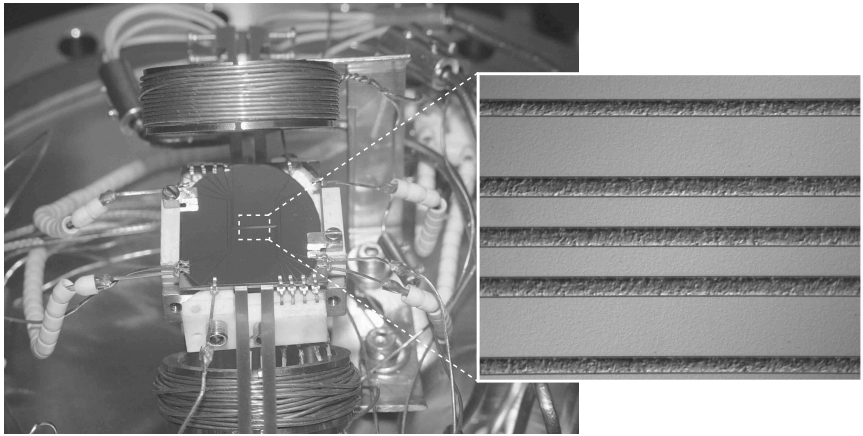,width=4.5in} \caption{Current
    carrying wire atom chip and chip mount used at Imperial College by
    Eriksson \textit{et al.} \cite{Eriksson_EPJD_2005}. The inset shows
    a high resolution microscope image of the central region of the chip
    where atoms are trapped and manipulated.}
    \label{fig:wire_chip}
\end{figure}

To give a concrete example of an atom chip, we show in
Fig.~\ref{fig:wire_chip} an interferometer
chip\cite{Eriksson_EPJD_2005} currently being used in our
laboratory. The reflective gold surface is used to form a mirror
magneto-optical trap (MOT) as a reservoir of pre-cooled atoms. The
coils used for the MOT are seen in the figure. The chip consists of
a 3\,$\mu$m layer of gold, thermally evaporated onto a silicon
substrate, in which wires have been lithographically defined by
ion-beam milling. Currents in the wires allow cold $^{87}$Rb atoms
to be magnetically trapped and split, in a matter-wave analogue of
an optical beam splitter \cite{Hinds_PRL_2001,Schumm_Nat_Phys_2005}.
The inset in Fig.~\ref{fig:wire_chip} is a microscope image showing
the four parallel wires at the heart of the interferometer, which
have a center-to-center distance of 300\,$\mu$m between the thick
outer pair.

With metallic chips such as this one, the homogeneity and stability
of cold atom clouds can be compromised close to the room temperature
surface. Two main phenomena have been identified: (i) spatial
imperfections of the wire cause the current to flow non-uniformly
and make the atom trap rough, leading to fragmentation of the atom
clouds \cite{Jones_JPB_2004,Moktadir_JMO_2007}, and (ii) thermal
fluctuations of the magnetic field near the surface, caused by
current noise in the conductor, drive spin flips of the atoms,
thereby inducing loss and destroying quantum coherences
\cite{Jones_PRL_2003}. The fragmentation can be controlled by using
fabrication methods that give very homogeneous wires with smooth
edges and uniform thickness. The magnetic noise field can be altered
by adjusting the thickness of the surface and by changing the
material in order to alter the complex conductivity
\cite{Rekdal_PRA_2004,Scheel_PRA_2005}.

We have also been using a chip based on patterns of permanent
magnetization, written on commercial videotape. This exhibits very
much longer spin flip times because the thermal noise currents are
suppressed by the high resistivity of the videotape. A pattern of
sinusoidal magnetisation (in plane), with a period of about
100\,$\mu$m, allows the confinement of atoms at distances less than
100\,$\mu$m from the surface of the chip in an array of long, thin
traps.  These have a very high aspect ratio ($>10^3$), with tight
transverse confinement that can bring ultra-cold atoms into the
one-dimensional regime \cite{Sinclair_PRA_2005,Sinclair_EPJD_2005}.
As well as offering a long spin-flip lifetime for BECs trapped near
its surface \cite{Sinclair_PRA_2005}, the permanent magnet atom chip
has the benefit of low power dissipation. Some fragmentation of
these videotape traps has been observed and is due primarily to
inhomogeneity of the magnetic layer. Preliminary experiments on
Pt/Co multilayer magnetic thin films \cite{Eriksson_APB_2004}
indicate that these are a promising alternative to videotape. Other
permanent magnet materials are being studied in the group of P.
Hannaford, also reporting at this confererence.

Superconductors offer another way to reduce the magnetic noise level
and increase the spin-flip lifetimes by many orders of magnitude.
Indeed, atoms have already been trapped near superconducting
surfaces in two laboratories
\cite{Haroche_PRL_2006,Shimizu_PRL_2007}. A recent
calculation\cite{Hohenester_PRA_2007} shows that rubidium atoms
trapped 1 $\mu$m away from a superconducting niobium surface at
liquid helium temperature should have a long spin flip lifetime, in
excess of 1000\,s. Under these conditions, the cold atoms offer a
new way to probe the superconducting surface because they are very
sensitive to the local magnetic field.  For example, cold atoms
would be able to image the vortices of a thin superconductor close
to the Kosterlitz-Thouless transition or more generally to study
vortex dynamics in a type II superconductor\cite{Scheel_PRA_2007}.

\section{Chips with optical micro-cavities}\label{section:microcavities}

\begin{figure}
\centering
    \psfig{file=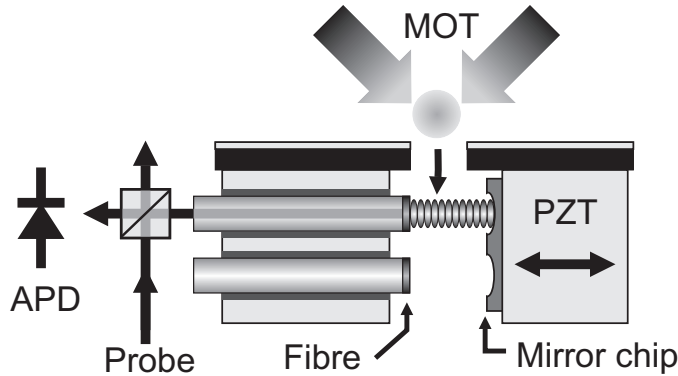,width=3.5in}
    \caption{Schematic diagram of
    apparatus used in the experiment.}
    \label{fig:apparatus}
\end{figure}

We have recently tested a new type of optical microcavity, which
combines the advantages of microfabrication and fibre-coupling with
the open access of a Fabry--Perot resonator. The plano-concave
resonator is formed between an isotropically etched hemisphere on
the surface of a silicon chip and the plane end of single-mode
fibre. Both surfaces are coated with a high-reflectivity multilayer
dielectric film. Our collaborators at the University of Southampton
have fabricated large arrays of concave mirrors on silicon chips and
we have used them to build cavities\cite{Trupke_APL_2005} with
$F>5000$ and $Q>10^6$. More recently, we have demonstrated both the
detection of atoms and the enhanced emission of photons into the
mode of the cavity.\cite{Trupke_PRL_2007} The apparatus is shown in
Fig.~\ref{fig:apparatus}. Two fibres with plane dielectric mirrors
on their ends are held in grooves on a glass-ceramic substrate,
facing two of the mirrors in the silicon array. The silicon mirror
chip is mounted on a piezo-electric translator (PZT), which is
adjusted to tune the upper cavity to the free-space atomic
resonance. In this work the lower cavity was not used. A mirror is
used to form a reflection MOT directly above the cavities.

For these experiments, we used cavities with $F=280$ and
$L=133\,\mu$m, giving $\kappa=2\pi\times 2\,$GHz.  The beam waist of
the mode is $w_\mathrm{C}=4.6\,\mu$m, giving a vacuum Rabi frequency
of $2g=2\pi\times 200$\,MHz for $^{85}$Rb atoms at an antinode of
the cavity mode, driven with circularly polarised light on the
closed transition \mbox{$|F=3,m_F=\pm
3\rangle\rightarrow|F'=4,m_{F'}=\pm 4\rangle$}. This transition has
a natural lifetime $2\gamma=2\pi\times 6$\,MHz, yielding a
single-atom cooperativity of $C=0.8$. The cooperativity increases
linearly with the number $N_\mathrm{A}$ of atoms in the cavity.
Since these are not generally at an antinode, the total
cooperativity is $C_{\mathrm{tot}}=
\tfrac{3}{7}C\sum_{n=1}^{N_\mathrm{A}}I(\mathbf{r}_n)$ where
$I(\mathbf{r}_n)$ is the fraction of peak intensity at the position
of the $n^{\mathrm{th}}$ atom. The factor $\tfrac{3}{7}$ accounts
for an average of Clebsch-Gordan coefficients over all the $F=3$
Zeeman sublevels. The effective number of atoms interacting with the
cavity field is then defined as $N_\mathrm{A}^{\mathrm{eff}}\equiv
C_{\mathrm{tot}}/(\tfrac{3}{7}C)$.

\subsection{Atom detection}
\begin{figure}
\centering
    \psfig{file=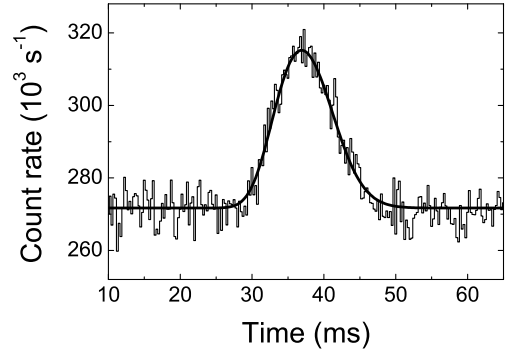,width=3.0in}
    \caption{Change in reflected
    probe light as atoms traverse the cavity. Laser, cavity, and atom frequencies are equal:
    $\omega_\mathrm{L}=\omega_\mathrm{C}=\omega_\mathrm{A}$. Average time-of-flight signal of 34 MOT releases, with an
    integration time of $250\,\mu$s. There are $\sim0.7$ atoms in
    the cavity on average at the peak.}
    \label{fig:TOF}
\end{figure}

To put atoms into the microcavity, we release a cloud of $\sim2
\times 10^7$ $^{85}$Rb atoms from the MOT, $7\,$mm above. Weak
resonant probe light ($\sim1$\,pW) is incident on the cavity through
a beam splitter and the reflected photons are counted by an
avalanche photodiode (APD, see Fig.~\ref{fig:apparatus}). When the
cavity is far from resonance with the light, the reflected intensity
is $I_\mathrm{max}=419\times10^3$\,s$^{-1}$, which drops to
$I_{\mathrm{min}}=272\times10^3$\,s$^{-1}$ at resonance. As
displayed in Fig.~\ref{fig:TOF}, this rises to a peak of
$I_{\mathrm{atoms}}=315\times10^3$\,s$^{-1}$ when the atoms pass
through the cavity. From these three count rates we derive the peak
cooperativity of the atom-cavity system using the relation
\begin{equation}
    \frac{I_{\mathrm{atoms}}}{I_{\mathrm{max}}}\simeq\left(-1+\frac{v}{P_\mathrm{tot}}\right)^2,
    \label{eq:reflIntensity}
\end{equation}
where $P_\mathrm{tot}=2C_\mathrm{tot}+1$ is an effective Purcell
factor, and $v=1-\sqrt{I_{\mathrm{min}}/I_{\mathrm{max}}}$ is the
empty cavity fringe visibility. From our experiments, we determine a
total cooperativity of $C_\mathrm{tot}\sim0.23$ at the peak of the
atom signal, corresponding to an effective atom number of only
$N_\mathrm{A}^{\mathrm{eff}}=0.7$.
\begin{figure}
\centering
    \psfig{file=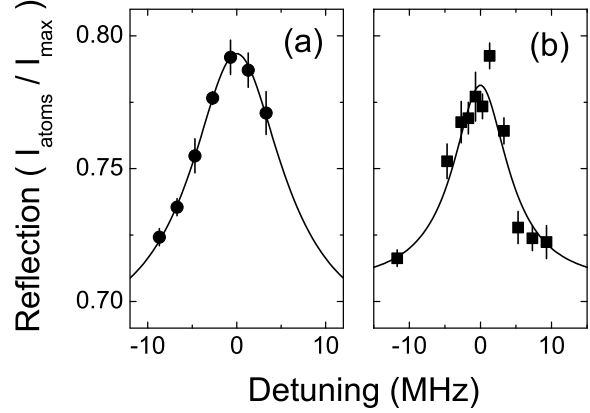,width=3.5in}
    \caption{Peak fraction of power reflected versus detuning $\Delta_\mathrm{LA}$
    when $\omega_\mathrm{C}=\omega_\mathrm{A}$ for two different initial MOT numbers.
    The fits (solid lines) yield average atom numbers of
    (a) $\langle N_\mathrm{A}^\mathrm{eff}\rangle=1.1$
    (b) $\langle N_\mathrm{A}^\mathrm{eff}\rangle=0.6$ \cite{Trupke_PRL_2007}.}
    \label{fig:detuning}
\end{figure}

We have measured the change in the peak reflected intensity when we
scan the detuning of the probe light
$\Delta_\mathrm{LA}=(\omega_\mathrm{L}-\omega_\mathrm{A})/\gamma$,
keeping the cavity frequency equal to the atomic resonance frequency
$\omega_\mathrm{C}=\omega_\mathrm{A}$. Figure~\ref{fig:detuning}
displays, for two different values of the initial atom number in the
MOT, $I_\mathrm{atoms}/I_\mathrm{max}$ versus $\Delta_\mathrm{LA}$,
for which we expect
\begin{equation}
    \frac{I_{\mathrm{atoms}}}{I_{\mathrm{max}}} \simeq
        \left|-1+\frac{v}{P_\mathrm{tot}}
            \frac{1+\Delta_\mathrm{LA}^2}{1+\Delta_\mathrm{LA}^2/P_\mathrm{tot}+2i\Delta_\mathrm{LA}C_\mathrm{tot}/P_\mathrm{tot}}\right|^2,
    \label{eq:reflDetuning}
\end{equation}
assuming $g/\gamma\gg(1,\Delta_\mathrm{LA})$. The solid lines in
Fig.~\ref{fig:detuning} are fits to the data, taking into account
fluctuations of $C_\mathrm{tot}$ and the probe laser linewidth
\cite{Trupke_PRL_2007}. These lineshapes confirm that we have better
than single atom sensitivity.

\subsection{Noise suppression}
\begin{figure}
\centering
    \psfig{file=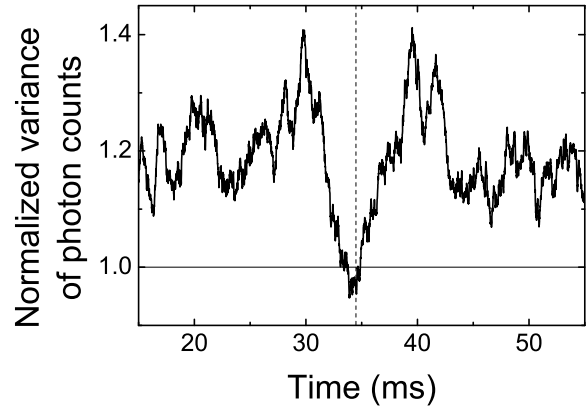,width=3.5in}
    \caption{Ratio of variance to mean for photon count rate reflected from the
    cavity during passage of atom cloud (corrected for APD dead-time and
    losses). Vertical dashed line: arrival time of the peak number of
    atoms. Horizontal line at 1.0: shot noise level.
    The sample size is 48 atom cloud drops.}
    \label{fig:noise}
\end{figure}

The arrival of atoms in the cavity is signaled not only by an
increase in the reflected light level, but also by a decrease in the
intensity noise. To measure this, we dropped the cloud 48 times,
recording the reflected intensity in $10\,\mu$s bins. We then
determined for each bin the ratio of variance to mean, taken over
the 48 drops. After correcting our count rate for the measured APD
dead-time of $44\,$ns \cite{Trupke_PRL_2007}, we found the correct
variance-to-mean ratio $f(n)\equiv Var(n)/\langle n\rangle$ at each
time bin ($n$ being the counts already corrected for the dead-time)
by including the $10\%$ loss at the beamsplitter of
Fig.~\ref{fig:apparatus} and the $60\%$ loss due to the
quantum-efficiency of the APD, according to
$f_\mathrm{corr}(n)-1=[f(n)-1]/0.54$. The resulting values, plotted
in Fig.~\ref{fig:noise}, show a strong reduction of the noise during
the passage of the atom cloud. In addition to the photon shot noise,
intensity fluctuations are due to small shifts of the cavity length
away from its resonance. The effect of these mechanical fluctuations
is expected to be less pronounced when atoms are present, and this
partially explains the decrease. However many of our experiments
seem to show additional noise reduction, the origin of which
requires further investigation.

\subsection{Photon generation}
\begin{figure}
\centering
    \psfig{file=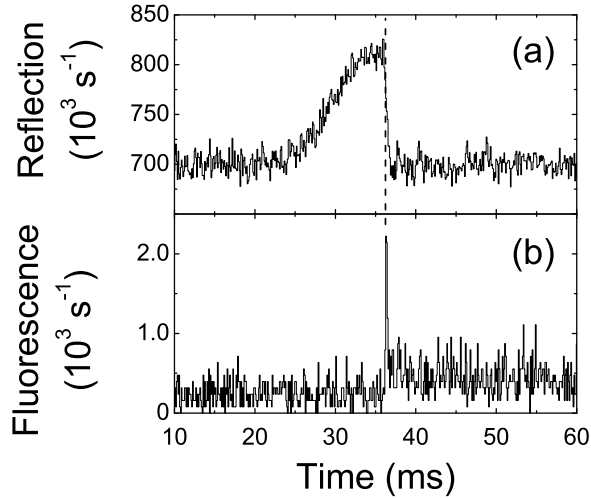,width=3.5in}
    \caption{(a) Sharp drop in the cavity reflection signal due to atom loss
    and optical pumping when the excitation laser is turned on. (b) Cavity-enhanced
    spontaneous emission collected via the fiber when the atom cloud is
    excited. The dotted line shows coincidence of the photon pulse with
    the turn-on of the excitation laser.}
    \label{fig:fluorescence}
\end{figure}

A further effect observed with our microcavity is the enhancement of
spontaneous emission.  For this experiment, we once again release a
cloud of atoms into the resonator. When the atom number in the
cavity reaches its highest value, we switch on a resonant excitation
beam, aimed at the atoms from below the cavity. This beam pushes the
atoms out of the cavity and also pumps them into the dark $F=2$
state. The loss of interacting atoms produces a sharp drop in the
cavity reflection signal, as shown in
Fig.~\ref{fig:fluorescence}~(a). We then perform the same experiment
but without any probe light. This time, a photon peak is recorded at
the moment the excitation laser is turned on, as shown in
Fig.~\ref{fig:fluorescence}~(b).  These are fluorescence photons
emitted into the cavity mode at the Purcell-enhanced rate
$4C_{\mathrm{tot}}\gamma$, as mentioned in the introduction, and
transported to the APD through the fibre. This is confirmed by the
fact that no photons are collected when the cavity is detuned from
the atomic transition or when there are no atoms in the cavity.

\section{Outlook}\label{section:outlook}

These experiments have shown that this type of microresonator is
suitable for the detection of numbers of atoms smaller than one, and
can be used to generate photons by enhanced spontaneous emission.
The latter is of particular interest for quantum-cryptography and
quantum-computation schemes. These cavities have by no means yet
reached their full potential. In the laboratory, with materials of
the same type, we have created cavities with $F>500$ and
$I_{\mathrm{min}}/I_{\mathrm{max}}<15\%$. Both values are
improvements by approximately a factor 2 compared to the published
results, which will considerably increase the detection signal, as
can be deduced from Eq.~(\ref{eq:reflIntensity}). We had previously
shown that we can even achieve finesse values of $F>5000$, albeit
with a low fringe-extinction of
$I_{\mathrm{min}}/I_{\mathrm{max}}\lesssim90\%$
\cite{Trupke_APL_2005}. At present, these values are limited by
scattering losses caused by the surface roughness of the silicon
mirror-substrate. Working with the group of M. Kraft at the
university of Southampton, we expect in the near future to have
smoother cavity mirrors that can operate well within the
strong-coupling regime of cavity quantum-electrodynamics, in which
the coherent exchange of quantum information between atoms and
photons in the cavity becomes possible.

These microcavities have the virtues that they are open, giving
atoms direct access to the peak of the mode pattern, they couple the
light in and out in a simple way through an integrated fiber, and
they can be fabricated in large numbers.  Being integrated into a
silicon substrate, they can be readily combined with other
micromachined components on atom chips. Some examples include the
magnetic trapping and guiding structures described in
Sec.~\ref{section:atom_chips}, which allow the deterministic
transport of atoms in and out of the cavity, integrated actuators
for tuning the resonator \cite{Gollasch_JMM_2005}, or pyramidal
micro-mirrors to realize an array a of single-atom sources from an
array of microscopic MOTs \cite{Trupke_APL_2006}. Work is in
progress to produce a fully integrated device above which neutral
atoms can be trapped, guided, detected and manipulated {\it in
situ}, using magnetic and optical fields.

\section*{Acknowledgments}

This work was supported by EU networks Atom Chips, Conquest, and
SCALA, by the Royal Society, and by EPSRC grants for QIPIRC, CCM
programme and Basic Technology. The atom chip devices in use at
Imperial College were fabricated by the nanosystems group of
Professor M. Kraft at the University of Southampton.

\bibliographystyle{ws-procs9x6}
\bibliography{proceedingsICOLS2007}

\end{document}